# On an identity for the volume integral of the square of a vector field

## A. M. Stewart

*Department of Theoretical Physics,
Research School of Physical Sciences and Engineering,
The Australian National University,
Canberra, ACT 0200, Australia.*

**ABSTRACT**
A proof is given of the vector identity proposed by Gubarev, Stodolsky and Zakarov that relates the volume integral of the square of a 3-vector field to non-local integrals of the curl and divergence of the field. The identity is applied to the case of the magnetic vector potential and magnetic field of a rotating charged shell. The latter provides a straightforward exercise in the use of the addition theorem of spherical harmonics.

**NOTE**
This ArXiv paper is based upon reference [9] but contains extra material. In section II, a derivation of equation (1) is given that is valid for the scalar product of two different vector fields and in Appendix B a proof that the general expression for the vector potential in the Coulomb gauge expressed in terms of the magnetic field[4,5] satisfies equation (2). In Appendix C, constraints on the electromagnetic gauge function are determined.

**I. INTRODUCTION**
Gubarev *et al.*[1] have reported an identity for a 3-vector field $\mathbf{A}(\mathbf{x})$ ($\mathbf{x} = x^1, x^2, x^3$) that is well behaved and vanishes sufficiently rapidly at spatial infinity. The identity is

$$\int d^3 x \, \mathbf{A}^2(\mathbf{x}) = \int d^3 x \int d^3 x' \frac{\nabla \cdot \mathbf{A}(\mathbf{x}) \nabla' \cdot \mathbf{A}(\mathbf{x}') + \nabla \times \mathbf{A}(\mathbf{x}) \cdot \nabla' \times \mathbf{A}(\mathbf{x}')}{4\pi |\mathbf{x} - \mathbf{x}'|} \quad . \quad (1)$$

The identity has obvious applications to electromagnetic theory. For example, it can be applied to the energy of the electromagnetic field or, if **A** is the magnetic vector potential in the Coulomb gauge with div**A** = 0, the first term on the right hand side of (1) vanishes and (1) can be expressed solely in terms of the magnetic field $\mathbf{B} = \text{curl}\mathbf{A}$

$$\int d^3 x \, \mathbf{A}^2(\mathbf{x}) = \int d^3 x \int d^3 x' \frac{\mathbf{B}(\mathbf{x}) \cdot \mathbf{B}(\mathbf{x}')}{4\pi |\mathbf{x} - \mathbf{x}'|} \quad . \quad (2)$$

Gubarev *et al.*[1] point out that (1) provides a gauge-invariant limitation on the nature of the magnetic vector potential, namely that, for a given magnetic field, (2) gives the minimum value for the volume integral of the square of the vector potential and this minimum value occurs in the Coulomb gauge. Gubarev *et al.*[1] used the generalization of the identity (1) to 4-vectors to investigate the topological structure of gauge field theories. Little further discussion has taken place on (1) as a 3-vector identity and it appears to warrant further examination.

Gubarev *et al.*[1] claim that the identity (1) can be proved by analogy with electrostatics and magnetostatics. However, this argument would miss any term, if it existed, that depended on both div**A** and curl**A** and vanished when either vanished. Alternatively, they claimed that the identity can be proved by extending to position space the identity $(\mathbf{k} \times \mathbf{A})^2 = \mathbf{k}^2 \mathbf{A}^2 - (\mathbf{k} \cdot \mathbf{A})^2$ in momentum space. However, this argument may be too cryptic for students beginning the study of electromagnetism.

In section II of this paper, we give a derivation of (1) which is based on the elementary principles of vector analysis and calculus and which should be more accessible to such students. Section III notes the form that the energy of the electromagnetic field takes when (1) is used. In section IV we demonstrate the application of (1) to a particular situation in electromagnetism, the rotating charged spherical shell.[2] This provides a straightforward exercise in the use of the addition theorem of spherical harmonics. In Appendix A a proof is given that each of the terms on the right hand side of (1) is greater than or equal to zero. In Appendix B, it is shown that the general expression for the vector potential in terms of the magnetic field[4,5] satisfies (2). In Appendix C, constraints on the electromagnetic gauge function are discussed.

**II. DERIVATION OF THE IDENTITY**

**A. Decomposition of the terms**

We start from the theorem of Helmholtz,[3,4,5,6] which states that an arbitrary vector field **A(x)** that is well behaved and vanishes sufficiently rapidly at infinity[11] may be expressed in the form

$$\begin{aligned}\mathbf{A}(\mathbf{x}) &= -\nabla f_a(\mathbf{x}) + \nabla \times \mathbf{F}_a(\mathbf{x}) \\ &= \mathbf{A}_l(\mathbf{x}) + \mathbf{A}_t(\mathbf{x})\end{aligned} \quad (3)$$

where the first and second terms are respectively known as the longitudinal (l) and transverse (t) terms. The scalar $f_a(\mathbf{x})$ and vector $\mathbf{F}_a(\mathbf{x})$ potentials are

$$f_a(\mathbf{x}) = \int d^3 x' \frac{\nabla' \cdot \mathbf{A}(\mathbf{x}')}{4\pi |\mathbf{x} - \mathbf{x}'|} \quad \text{and} \quad \mathbf{F}_a(\mathbf{x}) = \int d^3 x' \frac{\nabla' \times \mathbf{A}(\mathbf{x}')}{4\pi |\mathbf{x} - \mathbf{x}'|} \quad (4)$$

$\nabla = \hat{\mathbf{x}} \frac{\partial}{\partial x} + \hat{\mathbf{y}} \frac{\partial}{\partial y} + \hat{\mathbf{z}} \frac{\partial}{\partial z}$ is the usual vector differential operator with respect to **x** and $\nabla'$ the operator with respect to **x'**. Using (4), (1) becomes

$$\int d^3 x \, \mathbf{A}(\mathbf{x}) \cdot \mathbf{B}(\mathbf{x}) = \int d^3 x \, (-\nabla f_a + \nabla \times \mathbf{F}_a) \cdot (-\nabla f_b + \nabla \times \mathbf{F}_b) \quad , \quad (5)$$

following the suggestion of Durand[10] to extend the derivation to the scalar product of two different vector fields. Equation (5) gives three types of integral that need to be evaluated: products of differentials of the scalar and vector potentials and cross terms.

**B. Cross terms**

With the use of the vector identity

$$\nabla \cdot (f_a \nabla \times \mathbf{F}_b) = \nabla f_a \cdot (\nabla \times \mathbf{F}_b) \quad (6)$$

we find that the cross terms in (5) are zero because they become the integral of a total divergence which, using Gauss's theorem, turns into a surface integral that vanishes if the fields decay sufficiently rapidly at infinity.

**C. Longitudinal term**

The longitudinal term $\int d^3x\, \mathbf{A}_l(\mathbf{x})\cdot\mathbf{B}_l(\mathbf{x}) = \int d^3x\, \nabla f_a \cdot \nabla f_b$ is obtained by using the identity

$$\nabla\cdot(f_a \nabla f_b) = f_a \nabla^2 f_b + \nabla f_a \cdot \nabla f_b \qquad (7)$$

From

$$\nabla^2 f_b(\mathbf{x}) = \int d^3x'\, \frac{\nabla'\cdot\mathbf{B}(\mathbf{x}')}{4\pi}\nabla^2 \frac{1}{|\mathbf{x}-\mathbf{x}'|} \qquad (8)$$

together with the well-known identity[3,7]

$$\nabla^2 \frac{1}{|\mathbf{x}-\mathbf{x}'|} = -4\pi\delta(\mathbf{x}-\mathbf{x}') \qquad (9)$$

where δ is the Dirac delta function, we find

$$\nabla^2 f_b(\mathbf{x}) = -\nabla\cdot\mathbf{B}(\mathbf{x}) \qquad (10)$$

and hence, noting that the integral of the left hand side of (7) will give a vanishing surface integral, the integral of the first term on the right hand side of (7) comes to be

$$\int d^3x\, \mathbf{A}_l(\mathbf{x})\cdot\mathbf{B}_l(\mathbf{x}) = \int d^3x \int d^3x'\, \frac{\nabla'\cdot\mathbf{A}(\mathbf{x}')\nabla\cdot\mathbf{B}(\mathbf{x})}{4\pi|\mathbf{x}-\mathbf{x}'|} \qquad (11)$$

**D. Transverse term**

The integral of the transverse term is calculated with the help of the identity

$$\begin{aligned}\nabla\cdot[\mathbf{F}_a \times (\nabla\times\mathbf{F}_b)] &= (\nabla\times\mathbf{F}_a)\cdot(\nabla\times\mathbf{F}_b) - \mathbf{F}_a\cdot\nabla\times(\nabla\times\mathbf{F}_b) \\ &= (\nabla\times\mathbf{F}_a)\cdot(\nabla\times\mathbf{F}_b) - \mathbf{F}_a\cdot[\nabla(\nabla\cdot\mathbf{F}_b) - \nabla^2\mathbf{F}_b]\end{aligned} \qquad (12)$$

thereby giving, from the vanishing of the volume integral of the left hand side of (12),

$$\int d^3x\, \mathbf{A}_t(\mathbf{x})\cdot\mathbf{B}_t(\mathbf{x}) = \int d^3x\, \mathbf{F}_a\cdot[\nabla(\nabla\cdot\mathbf{F}_b) - \nabla^2\mathbf{F}_b] \qquad (13)$$

First we calculate div$\mathbf{F}_b$

$$\nabla\cdot\mathbf{F}_b(\mathbf{x}) = \int d^3x'\, \nabla\cdot\frac{\nabla'\times\mathbf{B}(\mathbf{x}')}{4\pi|\mathbf{x}-\mathbf{x}'|} \qquad (14)$$

From the identities

$$\nabla' \cdot \frac{\nabla' \times \mathbf{B}(\mathbf{x}')}{|\mathbf{x} - \mathbf{x}'|} = \nabla' \times \mathbf{B}(\mathbf{x}') \cdot \nabla' \frac{1}{|\mathbf{x} - \mathbf{x}'|} \quad \text{and} \quad \nabla \cdot \frac{\nabla' \times \mathbf{B}(\mathbf{x}')}{|\mathbf{x} - \mathbf{x}'|} = \nabla' \times \mathbf{B}(\mathbf{x}') \cdot \nabla \frac{1}{|\mathbf{x} - \mathbf{x}'|} \quad (15)$$

we find that

$$\nabla \cdot \frac{\nabla' \times \mathbf{B}(\mathbf{x}')}{|\mathbf{x} - \mathbf{x}'|} = -\nabla' \cdot \frac{\nabla' \times \mathbf{B}(\mathbf{x}')}{|\mathbf{x} - \mathbf{x}'|} \quad . \quad (16)$$

Accordingly, the integrand in (14) is a perfect divergence which turns into a vanishing surface integral. It follows that $\text{div}\mathbf{F}_b = 0$ and the first term on the right hand side of (13) vanishes.

With the help of (9) we find

$$\nabla^2 \mathbf{F}_b = \int d^3 x' \frac{\nabla' \times \mathbf{B}(\mathbf{x}')}{4\pi} \nabla^2 \frac{1}{|\mathbf{x} - \mathbf{x}'|} = -\nabla \times \mathbf{B}(\mathbf{x}) \quad . \quad (17)$$

Therefore the second term of (13) comes to

$$\int d^3 x \, \mathbf{A}_t(\mathbf{x}) \cdot \mathbf{B}_t(\mathbf{x}) |_2 = \int d^3 x \int d^3 x' \frac{\nabla' \times \mathbf{A}(\mathbf{x}') \cdot \nabla \times \mathbf{B}(\mathbf{x})}{4\pi |\mathbf{x} - \mathbf{x}'|} \quad (18)$$

giving

$$\int d^3 x \, \mathbf{A}(\mathbf{x}) \cdot \mathbf{B}(\mathbf{x}) = \int d^3 x \int d^3 x' \frac{\nabla' \cdot \mathbf{A}(\mathbf{x}) \nabla \cdot \mathbf{B}(\mathbf{x}) + \nabla' \times \mathbf{A}(\mathbf{x}') \cdot \nabla \times \mathbf{B}(\mathbf{x})}{4\pi |\mathbf{x} - \mathbf{x}'|} \quad (18a)$$

and (1) is proved. It is to be noted that since the Helmholtz theorem (3, 4) is valid for fields that depend on time[4,8,11] equations (1) and (18a) also apply to fields that depend on time.

### III. ENERGY OF THE ELECTROMAGNETIC FIELD
Using the Maxwell equations for div**E** and curl**E** the energy of the electric field (in SI units) is from (1)

$$U_E(t) = \frac{\varepsilon_0}{2} \int d^3 x \mathbf{E}^2 = \frac{1}{8\pi\varepsilon_0} \int d^3 x \int d^3 x' \left[ \frac{\rho(\mathbf{x},t)\rho(\mathbf{x}',t)}{|\mathbf{x}-\mathbf{x}'|} + \frac{\varepsilon_0^2 \partial \mathbf{B}(\mathbf{x},t)/\partial t \cdot \partial \mathbf{B}(\mathbf{x}',t)/\partial t}{|\mathbf{x}-\mathbf{x}'|} \right] \quad (19)$$

The first term describes the instantaneous energy of the charges present, while the second describes the energy in the electric field induced through Faraday's law when **B** changes with time. This generalizes the usual electrostatic result for the energy of a charge distribution to the time-dependent situation.

If we write as the total current **g** in terms of the charge transport current **j** and the displacement current

$$\mathbf{g}(\mathbf{x},t) = \mathbf{j}(\mathbf{x},t) + \varepsilon_0 \frac{\partial \mathbf{E}(\mathbf{x},t)}{\partial t} \quad (20)$$

and the inhomogeneous Maxwell equation in the form

$$\nabla \times \mathbf{B}(\mathbf{x},t) = \mu_0 \mathbf{g}(\mathbf{x},t) \quad , \tag{21}$$

the energy of the magnetic field may be expressed, using (1) as

$$U_M(t) = \frac{1}{2\mu_0}\int d^3x\, \mathbf{B}^2(\mathbf{x},t) = \frac{\mu_0}{8\pi}\int d^3x \int d^3x' \frac{\mathbf{g}(\mathbf{x},t)\cdot\mathbf{g}(\mathbf{x}',t)}{|\mathbf{x}-\mathbf{x}'|} \quad . \tag{22}$$

This has the same algebraic form as the energy of a static current distribution. There are no magnetic charges because div$\mathbf{B}$ = 0, so the term corresponding to the first in (19) is missing. From the appendix each of the terms of (19) is greater than or equal to zero as, of course, are both sides of (22).

### IV. APPLICATION TO THE ROTATING CHARGED SPHERICAL SHELL

**A. Potential and field**

We discuss a simple model by which the validity of the identity (1) can be verified in one case for the vector potential. The model we use is the charged spherical shell rotating about an axis through its center[2]. The shell is of radius $R$ with a charge $Q$ distributed over its surface uniformly. It rotates at an angular velocity $\omega$ about the $z$-axis.

The charge and current densities, using spherical coordinates $\mathbf{x} = (r, \theta, \phi)$, are

$$\rho(\mathbf{x}) = \frac{Q}{4\pi R^2}\delta(r-R) \quad \text{and} \quad \mathbf{j}(\mathbf{x}) = \frac{Q}{4\pi R^2}\delta(r-R)\boldsymbol{\omega}\times\mathbf{x} \tag{23}$$

and the magnetic moment of the shell is

$$\mathbf{m} = \int d^3x\, \mathbf{x}\times\mathbf{j}(\mathbf{x})/2 = \omega Q R^2/3 \quad . \tag{24}$$

The magnetic fields and potentials are given by de Castro[2]. For $0 < r < R$

$$\mathbf{A}(\mathbf{x}) = \frac{\mu_0 \mathbf{m}\times\mathbf{x}}{4\pi R^3} \quad \text{and} \quad \mathbf{B}(\mathbf{x}) = \frac{\mu_0 \mathbf{m}}{2\pi R^3} \tag{25}$$

and for $R < r < \infty$

$$\mathbf{A}(\mathbf{x}) = \frac{\mu_0 \mathbf{m}\times\mathbf{x}}{4\pi r^3} \quad \text{and} \quad \mathbf{B}(\mathbf{x}) = \frac{\mu_0}{4\pi}\frac{[3\hat{\mathbf{x}}(\mathbf{m}\cdot\hat{\mathbf{x}})-\mathbf{m}]}{r^3} \quad . \tag{26}$$

The magnetic vector potential $\mathbf{A}$ is continuous at $r = R$ and satisfies the condition $\nabla\cdot\mathbf{A} = 0$ everywhere. Inside the shell the magnetic field $\mathbf{B}$ is uniform. Outside the shell it has the form given by a magnetic dipole. The normal component of $\mathbf{B}$ is continuous at the surface of the shell.

**B. The volume integral**

First we calculate the volume integral of $\mathbf{A}^2$ in the region $0 < r < R$. From

$$(\mathbf{m} \times \mathbf{x}) \cdot (\mathbf{m} \times \mathbf{x}) = \mathbf{m}^2 r^2 - (\mathbf{m} \cdot \mathbf{x})^2 = \mathbf{m}^2 r^2 \sin^2 \theta \tag{27}$$

we get

$$\int_{r<R} d^3x\, \mathbf{A}^2(\mathbf{x}) = (\frac{\mu_0}{4\pi})^2 \int_0^{2\pi} d\phi \int_0^\pi \sin\theta\, d\theta \int_0^R r^2 dr\, \frac{\mathbf{m}^2 r^2 \sin^2\theta}{R^6} = \frac{(\mu_0 \mathbf{m})^2}{30\pi R} \quad, \tag{28}$$

using $\int_0^\pi \sin^3\theta\, d\theta = 4/3$.

For $R < r < \infty$ we find

$$\int_{R<r} d^3x\, \mathbf{A}^2(\mathbf{x}) = (\frac{\mu_0}{4\pi})^2 \int_0^{2\pi} d\phi \int_0^\pi \sin\theta\, d\theta \int_R^\infty r^2 dr\, \frac{\mathbf{m}^2 r^2 \sin^2\theta}{r^6} = \frac{(\mu_0 \mathbf{m})^2}{6\pi R} \tag{29}$$

Adding them together, we get for the volume integral of $\mathbf{A}^2$ over all space

$$\int d^3x\, \mathbf{A}^2(\mathbf{x}) = \frac{(\mu_0 \mathbf{m})^2}{5\pi R} \quad . \tag{30}$$

**C. The non-local integral**

We now evaluate the non-local integral on the right hand side of (2). We express the factor $1/|\mathbf{x} - \mathbf{x}'|$ in the form[7]

$$\frac{1}{|\mathbf{x} - \mathbf{x}'|} = \sum_{l=0}^\infty P_l(\cos\gamma) \frac{r_<^l}{r_>^{l+1}} \tag{31}$$

where $P_l$ is a Legendre polynomial, $r_<$ and $r_>$ are the lesser and greater parts of the magnitudes of $\mathbf{x}$ and $\mathbf{x}'$ and $\gamma$ is the angle between those two vectors given by $\cos\gamma = \cos\theta\cos\theta' + \sin\theta\sin\theta'\cos(\phi - \phi')$. Invoking the addition theorem of spherical harmonics $Y_{lm}(\theta,\phi)$ [7]

$$P_l(\cos\gamma) = \frac{4\pi}{2l+1} \sum_{m=-l}^{+l} Y_{lm}^*(\theta',\phi') Y_{lm}(\theta,\phi) \quad . \tag{32}$$

gives $\quad \dfrac{1}{|\mathbf{x} - \mathbf{x}'|} = \sum_{l=0}^\infty \sum_{m=-l}^{+l} \dfrac{4\pi}{2l+1} \dfrac{r_<^l}{r_>^{l+1}} Y_{lm}^*(\theta',\phi') Y_{lm}(\theta,\phi) \tag{33}$

where the orthogonality relationship for spherical harmonics is

$$\int d\Omega\, Y_{l'm'}^*(\theta,\phi) Y_{lm}(\theta,\phi) = \delta_{l',l}\delta_{m',m} \tag{34}$$

$\Omega$ being the solid angle corresponding to $\theta,\phi$. In the cases where the terms that we integrate do

not depend on $\phi$, we deduce that only the $m = 0$ terms of (33) will survive the angular integrations and (33) becomes

$$\frac{1}{|\mathbf{x} - \mathbf{x}'|} => \sum_{l=0}^{\infty} \frac{4\pi}{2l+1} \frac{r_<^l}{r_>^{l+1}} Y_{l0}(\theta') Y_{l0}(\theta) \tag{35}$$

The double integration of the right hand side of (2) is composed of three parts: inside-inside, outside-outside and inside-outside. The first part is the simplest because the magnetic field is uniform over the volume inside the sphere.

*1. Inside-inside*
The integral of the right-hand side of (2) for the inside-inside region is

$$\frac{\mathbf{B}^2}{4\pi} \sum_{l=0}^{\infty} \int d\Omega \, Y_{l0}(\theta) \int d\Omega' \, Y_{l0}(\theta') \int_0^R dr \, r^2 \int_0^R dr' \, r'^2 \frac{4\pi}{2l+1} \frac{r_<^l}{r_>^{l+1}} \tag{36}$$

where the volume element $d^3x$ is given by $d^3x = d\Omega r^2 dr$. Unity is replaced by $2\pi^{1/2} Y_{00}$[7] in the angular integrals and, from the orthogonality of the spherical harmonics (34), the integral vanishes unless $l = 0$. Expression (36) then becomes

$$4\pi \mathbf{B}^2 \int_0^R dr \, r^2 \int_0^R dr' \, r'^2 \frac{1}{r_>} \tag{37}$$

or

$$4\pi \mathbf{B}^2 \int_0^R dr \, r^2 [\int_0^r dr' \, r'^2 \frac{1}{r} + \int_r^R dr' \, r'^2 \frac{1}{r'}] = 4\pi \mathbf{B}^2 \frac{2R^5}{15} \tag{38}$$

where the first integral is for the region $0 < r' < r$ and the second for $r < r' < R$. The integral in (38) is easily evaluated to give $8\pi \mathbf{B}^2 R^5/15$ or $2(\mu_0 \mathbf{m})^2/R 15\pi$.

*2. Inside-outside*
The integral of the right-hand side of (2) for the inside-outside region contains, from (25) and (26), terms that depend on $\theta$ and $\theta'$ in the form with $0 < r' < R$ and $R < r < \infty$

$$\mathbf{B}(\mathbf{x}) \cdot \mathbf{B}(\mathbf{x}') = 2(\frac{\mu_0}{4\pi})^2 \frac{3(\mathbf{m} \cdot \hat{\mathbf{x}}) - \mathbf{m}^2}{R^3 r^3}$$

$$= 2\left(\frac{\mu_0 \mathbf{m}}{4\pi}\right)^2 \frac{3\cos^2\theta - 1}{R^3 r^3} \tag{39}$$

Noting that the above expression is proportional in its angular variation to the spherical harmonic $Y_{20} = \frac{1}{4}\sqrt{\frac{5}{\pi}}(3\cos^2\theta - 1)$, the integrals over the angles are proportional to

$$\sum_{l,m}^{\infty} \int d\Omega \, Y_{lm}(\theta,\phi) Y_{20}(\theta) \int d\Omega' \, Y_{lm}(\theta',\phi') \tag{40}$$

From the orthogonality of the spherical harmonics it is required that $l = 2$ from the $\theta$ integral and $l = 0$ from the $\theta'$ integral for the integral to be non-vanishing. It is impossible to satisfy these conditions simultaneously so the inside-outside integral vanishes.

### *3. Outside-outside*

The integral of the right hand side of (2) for the outside-outside region contains, from (26), terms that depend on $\theta$ and $\theta'$ in the form

$$\mathbf{B}(\mathbf{x}) \cdot \mathbf{B}(\mathbf{x}') = (\frac{\mu_0}{4\pi})^2 \frac{\mathbf{m}^2 - 3(\mathbf{m} \cdot \hat{\mathbf{x}})^2 - 3(\mathbf{m} \cdot \hat{\mathbf{x}}')^2 + 9\hat{\mathbf{x}} \cdot \hat{\mathbf{x}}'(\mathbf{m} \cdot \hat{\mathbf{x}})(\mathbf{m} \cdot \hat{\mathbf{x}}')}{r^3 r'^3}$$

$$= \left(\frac{\mu_0 \mathbf{m}}{4\pi}\right)^2 \frac{1 - 3(\cos^2\theta + \cos^2\theta') + 9\cos\gamma \cos\theta \cos\theta'}{r^3 r'^3} \qquad (41)$$

It can be verified, for example by using a program that does algebra on the computer, that the numerator of (41) can be expressed as the finite sum of products of spherical harmonics

$$1 - 3(\cos^2\theta + \cos^2\theta') + 9\cos\gamma\cos\theta\cos\theta' =$$
$$\frac{16\pi}{5}\{Y_{20}(\theta)Y_{20}(\theta') + \frac{3}{4}[Y^*_{21}(\theta,\phi)Y_{21}(\theta',\phi') + Y^*_{2-1}(\theta,\phi)Y_{2-1}(\theta',\phi')]\} \qquad (42)$$

where $Y_{2\pm1}(\theta,\phi) = \mp\sqrt{\frac{15}{8\pi}}\cos\theta\sin\theta e^{\pm i\phi}$. When the orthogonality relation (34) is used, the integrals over the angles of the spherical harmonics in (33) and (42) produce a factor of $8\pi$. The integration for this region therefore becomes

$$\frac{8\pi}{5}\left(\frac{\mu_0 \mathbf{m}}{4\pi}\right)^2 \int_R^\infty dr\, r^2 \int_R^\infty dr'\, r'^2 \frac{1}{r^3 r'^3} \frac{r_<^2}{r_>^3} \qquad (43)$$

with the $(2l+1)$ factor in (33) for $l=2$ giving rise to a factor of 5 in the denominator. Writing out the integrations over $r$ and $r'$ explicitly we get

$$\frac{(\mu_0\mathbf{m})^2}{10\pi}\int_R^\infty \frac{dr}{r}[\int_R^r \frac{dr'}{r'}\frac{r'^2}{r^3} + \int_r^\infty \frac{dr'}{r'}\frac{r^2}{r'^3}] = \frac{(\mu_0\mathbf{m})^2}{10\pi}\frac{2}{3R} = \frac{(\mu_0\mathbf{m})^2}{15\pi R} \qquad (44)$$

where the first integral is for the region $R < r' < r$ and the second for $r < r' < \infty$. Adding the result of (44) to the inside-inside value of $2(\mu_0\mathbf{m})^2/15\pi R$ we get a total contribution of $(\mu_0\mathbf{m})^2/5\pi R$. This is in agreement with the value obtained from the volume integral of $\mathbf{A}^2$. The identity of Gubarev *et al.* has therefore been verified for the case of the rotating charged shell in the Coulomb gauge.

### V. CONCLUSIONS

A new identity for 3-vector fields reported by Gubarev *et al.*[1] has been reviewed and a formal proof of an extension of it has been given using elementary principles of vector analysis. The identity is of interest when applied to electromagnetism as it shows how, for a given distribution of magnetic field, the volume integral of the square of the magnetic vector potential has a minimum value when the Coulomb gauge condition div$\mathbf{A} = 0$ holds. In this sense the Coulomb gauge can be said to be a minimal gauge.

The identity may be applied to the energy of the electromagnetic field and the identity has been explicitly verified for a simple model, the rotating charged spherical shell, for which the electromagnetic vector potential and magnetic field are both known.[2]

**APPENDIX A: PROOF OF AN INEQUALITY**

In order to show that both terms in (1) are greater than or equal to zero we need to show that

$$I = \int d^3x \int d^3x' \frac{f(\mathbf{x})f(\mathbf{x}')}{|\mathbf{x} - \mathbf{x}'|} \geq 0 \tag{A1}$$

for any well behaved scalar function $f(\mathbf{x})$. This may done by expressing by substituting into (A1) the interaction term in the form[7]

$$\frac{1}{|\mathbf{x} - \mathbf{x}'|} = \frac{1}{2\pi^2} \int d^3q \frac{e^{i\mathbf{q}\cdot(\mathbf{x} - \mathbf{x}')}}{\mathbf{q}^2} \tag{A2}$$

to get

$$I = \frac{1}{2\pi^2} \int d^3q \frac{1}{\mathbf{q}^2} \int d^3x \int d^3x' f(\mathbf{x}) f(\mathbf{x}') e^{i\mathbf{q}\cdot(\mathbf{x} - \mathbf{x}')} . \tag{A3}$$

This can be seen to give

$$I = 4\pi \int d^3q \frac{h(\mathbf{q})h(-\mathbf{q})}{q^2} \tag{A4}$$

where $h(\mathbf{q})$ is the Fourier transform of $f(\mathbf{x})$

$$h(\mathbf{q}) = \frac{1}{(2\pi)^{3/2}} \int d^3x \, e^{-i\mathbf{q}\cdot\mathbf{x}} f(\mathbf{x}) . \tag{A5}$$

If $f(\mathbf{x})$ is real then $h(-\mathbf{q}) = h^*(\mathbf{q})$, so from (A4) $I$ is greater than or equal to zero as required. The integral (A4) will remain finite if $h(\mathbf{q})$ vanishes as $\mathbf{q}$ approaches infinity.

**APPENDIX B: INTEGRAL OF THE SQUARE OF A GENERAL POTENTIAL**

The formal solution to the equation

$$\mathbf{B}(\mathbf{x},t) = \text{curl}\mathbf{A}(\mathbf{x},t) \tag{B1}$$

where **B** is a divergenceless (div**B** = 0) vector field is[4,5]

$$\mathbf{A}(\mathbf{x},t) = \nabla_x \times \frac{1}{4\pi} \int d^3y \frac{\mathbf{B}(\mathbf{y},t)}{|\mathbf{x} - \mathbf{y}|} + \lambda \nabla_x \chi(\mathbf{x},t) \tag{B2}$$

where the gradient is taken with respect to **x**, $\lambda$ is a number and $\chi(\mathbf{x},t)$ is any well-behaved scalar field. The term containing λ is the pure gauge term. The divergence of **A** is given by div**A** = $\lambda \nabla^2 \chi(\mathbf{x},t)$ so for λ = 0 the Coulomb gauge is obtained. We take λ = 0, suppress the time variable and calculate the integral over all space of the square of the first term of (B2), the Coulomb gauge term, to confirm that it satisfies (2):

$$\int d^3x\, \mathbf{A}^2(\mathbf{x}) = \frac{1}{(4\pi)^2}\int d^3x \int d^3y \int d^3z\, \nabla_x \times \frac{\mathbf{B}(\mathbf{y})}{|\mathbf{x}-\mathbf{y}|} \cdot \nabla_x \times \frac{\mathbf{B}(\mathbf{z})}{|\mathbf{x}-\mathbf{z}|} \quad . \tag{B3}$$

Using the vector identity for $\nabla \cdot (\mathbf{C} \times \mathbf{D})$ we obtain the relation

$$\nabla_x \cdot \{[\nabla_x \times \frac{\mathbf{B}(\mathbf{y})}{|\mathbf{x}-\mathbf{y}|}] \times [\frac{\mathbf{B}(\mathbf{z})}{|\mathbf{x}-\mathbf{z}|}]\} = \frac{\mathbf{B}(\mathbf{z})}{|\mathbf{x}-\mathbf{z}|} \cdot \nabla_x \times \nabla_x \times \frac{\mathbf{B}(\mathbf{y})}{|\mathbf{x}-\mathbf{y}|} - \nabla_x \times \frac{\mathbf{B}(\mathbf{y})}{|\mathbf{x}-\mathbf{y}|} \cdot \nabla_x \times \frac{\mathbf{B}(\mathbf{z})}{|\mathbf{x}-\mathbf{z}|} \tag{B4}$$

so, noting that the integral of the left hand side vanishes by Gauss' theorem, we find

$$\int d^3x\, \mathbf{A}^2(\mathbf{x}) = \frac{1}{(4\pi)^2}\int d^3x \int d^3y \int d^3z\, \frac{\mathbf{B}(\mathbf{z})}{|\mathbf{x}-\mathbf{z}|} \cdot \nabla_x \times \nabla_x \times \frac{\mathbf{B}(\mathbf{y})}{|\mathbf{x}-\mathbf{y}|} \quad . \tag{B5}$$

It can be shown, using the identity for the curl of the curl of a vector field, the identity

$$\nabla_x [\mathbf{B}(\mathbf{y}) \cdot \nabla_x \frac{1}{|\mathbf{x}-\mathbf{y}|}] = [\mathbf{B}(\mathbf{y}) \cdot \nabla_x] \nabla_x \frac{1}{|\mathbf{x}-\mathbf{y}|} \tag{B6}$$

and (9), that

$$\nabla_x \times \nabla_x \times \frac{\mathbf{B}(\mathbf{y})}{|\mathbf{x}-\mathbf{y}|} = 4\pi \mathbf{B}(\mathbf{y})\delta(\mathbf{x}-\mathbf{y}) + [\mathbf{B}(\mathbf{y}) \cdot \nabla_x]\nabla_x \frac{1}{|\mathbf{x}-\mathbf{y}|} \quad . \tag{B7}$$

Substituting the second term of (B7) into (B5) we get

$$\frac{1}{(4\pi)^2}\int d^3x \int d^3y \int d^3z\, \frac{\mathbf{B}(\mathbf{z})}{|\mathbf{x}-\mathbf{z}|} \cdot [\mathbf{B}(\mathbf{y}) \cdot \nabla_x]\nabla_x \frac{1}{|\mathbf{x}-\mathbf{y}|} \quad . \tag{B8}$$

Using the identity for the divergence of a vector field times a scalar field we get

$$\nabla_x \cdot \{[\frac{\mathbf{B}(\mathbf{z})}{|\mathbf{x}-\mathbf{z}|}][\mathbf{B}(\mathbf{y}) \cdot \nabla_x \frac{1}{|\mathbf{x}-\mathbf{y}|}]\} = $$
$$\nabla_x \cdot \{\frac{\mathbf{B}(\mathbf{z})}{|\mathbf{x}-\mathbf{z}|}\}\mathbf{B}(\mathbf{y}) \cdot \nabla_x \frac{1}{|\mathbf{x}-\mathbf{y}|} + \frac{\mathbf{B}(\mathbf{z})}{|\mathbf{x}-\mathbf{z}|} \cdot \nabla_x \{\mathbf{B}(\mathbf{y}) \cdot \nabla_x \frac{1}{|\mathbf{x}-\mathbf{y}|}\} \tag{B9}$$

With the identity for the gradient of the scalar product of two vectors the right hand side of (B9) comes to

$$\nabla_x \cdot \{\frac{\mathbf{B}(\mathbf{z})}{|\mathbf{x}-\mathbf{z}|}\}\nabla_x \cdot \{\frac{\mathbf{B}(\mathbf{y})}{|\mathbf{x}-\mathbf{y}|}\} + \frac{\mathbf{B}(\mathbf{z})}{|\mathbf{x}-\mathbf{z}|} \cdot [\mathbf{B}(\mathbf{y}) \cdot \nabla_x]\nabla_x \frac{1}{|\mathbf{x}-\mathbf{y}|}\} \quad . \tag{B10}$$

By switching the arguments of the gradients we find for the second term that arises from (B7)

$$-\frac{1}{(4\pi)^2}\int d^3x \int d^3y \nabla_y \cdot \frac{\mathbf{B(y)}}{|\mathbf{x-y}|} \int d^3z \nabla_z \cdot \frac{\mathbf{B(z)}}{|\mathbf{x-z}|} \qquad . \qquad (B11)$$

Since the *y* and *z* integrands are perfect differentials this term vanishes. We are left with the first term of (B7) which gives the required result

$$\int d^3x\, \mathbf{A}_c^2(\mathbf{x},t) = \int d^3x \int d^3x' \frac{\mathbf{B(x},t)\cdot\mathbf{B(x'},t)}{4\pi|\mathbf{x-x'}|} \qquad . \qquad (B12)$$

**APPENDIX C: CONSTRAINTS ON THE ELECTROMAGNETIC GAUGE FUNCTION**

We find the conditions that the electromagnetic gauge function $\chi(\mathbf{x},t)$ must satisfy for (1) to give a minimum for the Coulomb gauge ($\nabla \cdot \mathbf{A}_c = 0$). We make a gauge transformation from the Coulomb gauge vector potential to another gauge $\mathbf{A}_c \rightarrow \mathbf{A}' = \mathbf{A}_c + \nabla\chi$ so that

$$\int d^3x\, \mathbf{A}'^2 = \int d^3x\, \mathbf{A}_c^2 + \int d^3x\, \nabla\chi \cdot \nabla\chi + 2\int d^3x\, \nabla\chi \cdot \mathbf{A}_c \qquad . \qquad (C1)$$

For the minimum property to hold it is necessary that the last term of (C1) vanishes. Consider the general vector identity

$$\nabla \cdot (\chi \mathbf{A}) = \chi \nabla \cdot \mathbf{A} + \mathbf{A} \cdot \nabla \chi \qquad . \qquad (C2)$$

and integrate over all space. The first term on the right of (C2) is zero because we use the Coulomb gauge, and the integral of the left hand side gives, by Gauss' theorem, the surface integral at $r \rightarrow \infty$, which is required to vanish

$$\int d\mathbf{S} \cdot \chi \mathbf{A}_c = \int d\Omega\, r^2 \chi A_c^r \qquad . \qquad (C3)$$

Spherical coordinates $\{r, \theta, \phi\}$ are used; $A_c^r$ is the radial component of the vector potential, and $\Omega$ is the solid angle. A spherical surface of integration is used.

The vector potentials of electromagnetic fields in the Coulomb gauge have been obtained in another paper.[11] For the static field of a magnetic moment *m* directed in the *z*-direction it is

$$\mathbf{A}_C(\mathbf{r}) = (\mu_0 m / 4\pi r^2)\{0, 0, \sin\theta\} \qquad . \qquad (C4)$$

For electric dipole radiation due to an electric dipole *p* vibrating with angular frequency $\omega$ in the *z*-direction it is

$$\mathbf{A}_C(\mathbf{r},t) = (\frac{p\mu_0}{4\pi})(\frac{\omega}{r})\sin[\omega(t - r/c)]\{0, \sin\theta, 0\} \qquad . \qquad (C5)$$

This expression is usually derived in the Lorenz gauge;[7] the Coulomb gauge version, which is given here, has been obtained elsewhere.[11] The potential for an oscillating magnetic dipole *m* directed in the *z*-direction has leading term

$$\mathbf{A}_C(\mathbf{r},t) = -(\omega/c)(\frac{m\mu_0}{4\pi})\frac{\sin\theta}{r}\sin[\omega(t-r/c)]\{0,0,1\} \qquad . \qquad (C6)$$

This expression is valid in both the Lorenz and the Coulomb gauges.[11]

It is seen that in all cases the radial component of the vector potential is zero. It contributes nothing to (C3). Accordingly, for the last term of (C1) to vanish and for that equation to represent a minimum principle, the only requirement for the behaviour of the electromagnetic gauge function χ is that it should not diverge at spatial infinity. It need not vanish there; it could be uniform or oscillatory.

## ACKNOWLEDGEMENTS

K. Kumar and D.A. Woodside are thanked for helpful comments. The referees are thanked for suggestions for improving the clarity of the paper and the Australian Defence Force Academy is thanked for providing library facilities.